\documentclass[prd,preprintnumbers,floatfix,aps,nofootinbib,notitlepage]{revtex4}

\pdfoutput=1

\usepackage{latexsym}
\usepackage{epsfig}
\usepackage{amssymb}
\usepackage{amsmath}
\usepackage{color}
\usepackage{mathtools}
\usepackage{hhline}

\newcommand{\lp}{\left(}
\newcommand{\rp}{\right)}
\newcommand{\lb}{\left[}
\newcommand{\rb}{\right]}

\newcommand{\ba}{\begin{eqnarray}}
\newcommand{\ea}{\end{eqnarray}}
\newcommand{\be}{\begin{equation}}
\newcommand{\ee}{\end{equation}}

\newcommand{\al}{\alpha}
\newcommand{\bt}{\beta}
\newcommand{\ga}{\gamma}

\newcommand{\da}{\delta}
\newcommand{\la}{\lambda}

\newcommand{\en}{\epsilon}
\newcommand{\ve}{\varepsilon}

\newcommand{\Sa}{\Sigma}

\newcommand{\h}{h}

\newcommand{\p}{p}

\newcommand{\cM}{\mathcal{M}}
\newcommand{\e}{e^{-2\al+4\sigma}}
\newcommand{\rar}{\rightarrow}

\newcommand{\tg}{\tilde{\gamma}}

\newcommand{\lK}{\lambda_\h}
\newcommand{\lM}{\lambda_M}
\newcommand{\lV}{\lambda_V}

\newcommand{\vx}{{\vec x}}

\newcommand{\SQRT}{\sqrt{\star}}

\begin{document}

\title{Disformal vectors and anisotropies on a warped brane\\\vspace*{5pt}\footnotesize{Hulluilla on Halvat Huvit}}

\author{Tomi S. Koivisto}
\email{tomik@astro.uio.no}
\affiliation{Nordita, KTH Royal Institute of Technology and Stockholm University, Roslagstullsbacken 23, SE-10691 Stockholm, Sweden}

\author{Federico R. Urban}
\email{furban@ulb.ac.be}
\affiliation{Service de Physique Th\'eorique, Universit\'e Libre de Bruxelles, CP225, Boulevard du Triomphe, B-1050 Brussels, Belgium}

\date{\today}

\begin{abstract}

The Maxwell action is conformally invariant and classically ignorant of conformally flat metrics. However, if the vector lives in a disformal metric--- as it does if residing upon a moving brane---this is no longer true. The disformal coupling is then mediated by a Dirac-Born-Infeld scalar field. Here a systematic dynamical system analysis is developed for anisotropic Bianchi I cosmology with a massive disformally coupled vector field. Several new fixed points are found, including anisotropic scaling solutions. The presented formalism here presented can be conveniently applied to general scenarios with or without extra dimensional motivations. This is illustrated here by performing a complete analysis with simple assumption that both the potentials and the warp factor for the brane are (nearly) exponential. In that case, the anisotropic fixed points are either not attractors, do not describe accelerating expansion or else they feature too large anisotropies to be compatible with observations. Nonetheless, viable classes of models exist where isotropy is retained due to rapid oscillations of the vector field, thus providing a possible realisation of disformally interacting massive dark matter.

\end{abstract}

\preprint{NORDITA-2014-87}

\maketitle


\section{Introduction}

Higher dimensional theories predict couplings of scalar and vector fields. Already the almost century-old Kaluza-Klein model featured a scalar-Maxwell coupling $\phi^2 F^2$. Such a coupling is present in the bosonic sector of supergravity and can be obtained from various versions of string theories. However, an interaction of different nature emerges generically in brane-world scenarios where matter resides on a moving brane \cite{KOIVISTO:2013jwa,Koivisto:2013fta}: this is due to the {\it disformal} relation of the induced metric upon the brane and the gravitational metric in our four dimensions. In string theory, the D-branes are identified as boundary conditions for the endpoints of the strings described by $U(1)$ gauge fields; thus it appears very natural to take into account matter on the brane in the form of vector fields \cite{Dimopoulos:2011pe,Wills:2013yga}.

Even though to much lesser extent than the $\phi^2 F^2$ class of theories \cite{EspositoFarese:2009aj}, disformally coupled vector theories have been explored in various contexts in cosmology. The property of disformally related metrics having different causal structure has been exploited in bimetric varying speed of light theories that could provide an alternative to cosmological inflation \cite{Magueijo:2003gj}. The disformal relation is also necessary in relativistic MOND theories to simulate the bending of light caused by dark matter \cite{Skordis:2009bf}. Quite recently the implications of the coupling in the electromagnetic sector were investigated and interesting phenomenology \footnote{In the case of nonrelativistic matter, the coupling is typically screened from local experiments \cite{Koivisto:2012za} (see  \cite{Burrage:2014uwa} on screening the conformal part of the coupling in a DBI setting).} was uncovered at the level of laboratory experiments and cosmology \cite{Brax:2012ie,vandeBruck:2013yxa,Brax:2013nsa}. A disformal generalisation of the conformal invariance of the Maxwell action was found in \cite{Goulart:2013laa}.

In this paper our purpose is to systematically study the cosmological dynamics of a disformally coupled vector field. We define a generic class of models that could be explicitly derived from the higher-dimensional set-up wherein the vector field resides on a moving brane. The scalar field has then a Dirac-Born-Infeld (DBI) action and the precise form of the coupling to the vector field is determined by the warped extra-dimensional geometry. For generality, we also allow a mass term for the vector. This breaks the gauge invariance of the Maxwell theory, and, consequently, the mass can also become a function of the conformal part of the coupling. We then end up with a new well-motivated scalar-vector theory and take the first steps in exploring its potential for cosmology.

One relevant application of this theory would be vector field models of inflation and dark energy \cite{Koivisto:2008xf}. Although vectors have been studied in the context of inflationary magnetogenesis \cite{Turner:1987bw}, their possible role as drivers of inflation \cite{Ford:1989me} was not actively considered until recently. In the model proposed in \cite{Golovnev:2008cf,Golovnev:2008hv}, a specific non-minimal coupling of the vector field facilitates its slow rolling. The model as such may present some problematic aspects, in particular the appearance of anisotropies and instabilities in perturbations \cite{Himmetoglu:2008hx,Pitrou:2008gk,Golovnev:2009ks,Golovnev:2011yc}, however, variations of the model such as its generalisation to and with other p-forms have been explored \cite{Koivisto:2009sd,Germani:2009iq}. With three-forms (i.e., vector fields with noncanonical kinetic terms) both isotropy and slow roll can be realised quite naturally \cite{Koivisto:2009ew,Koivisto:2009fb,Boehmer:2011tp}, and three-form cosmologies have been found to feature viable inflation \cite{DeFelice:2012jt,Mulryne:2012ax,Kumar:2014oka}, reheating \cite{DeFelice:2012wy}, a new mechanism for magnetogenesis \cite{Koivisto:2011rm,Urban:2012ib} (that however may produce excessive anisotropies \cite{Urban:2012ib}) and novel coupled dark energy cosmologies \cite{Ngampitipan:2011se,Koivisto:2012xm}.

An interesting possibility is that of vector fields supporting anisotropy during inflation under the guise of ``anisotropic hair'' \cite{Kanno:2008gn,Watanabe:2009ct,Hervik:2011xm} which the vectors themselves generate and sustain. Without a mechanism of this sort, one needs to finely tune the initial conditions for inflation around 60 e-folds before its beginning in order to obtain anisotropic signatures at the largest scales, of which there appears to be some evidence in the cosmic microwave background radiation (CMB), see \cite{Copi:2013jna,Ade:2013zuv} and references to earlier works therein. Such possibilities have motivated the development of perturbation theory in anisotropic cosmologies \cite{Pereira:2007yy,Zlosnik:2011iu,Gumrukcuoglu:2007bx,Emami:2014tpa}. In general, the presence of vector fields often translates into the appearance of observable (possibly anisotropic) signatures in the statistical correlators of cosmological perturbations \cite{Dimopoulos:2008yv,Karciauskas:2008bc,Chang:2013mya,Chen:2014eua}. Indeed, if not responsible for the dynamics of the background, vectors could still act as a curvaton generating the primordial perturbations \cite{Dimopoulos:2006ms,Dimopoulos:2009am,Dimopoulos:2009vu,Namba:2012gg}. 

Vectors have been also employed as a field theoretical realisation of anisotropic dark energy \cite{Koivisto:2007bp,Koivisto:2008ig}, (see also \cite{ArmendarizPicon:2004pm,Akarsu:2009gz,Kumar:2011rn,Thorsrud:2012mu,Akarsu:2013dva}). Viable generalisations of the models include Horndeski actions for the vector field \cite{Barrow:2012ay,Jimenez:2013qsa} and nonmetric theories where the field can be identified with the Weyl vector describing the (nonmetric) property of the spacetime connection \cite{Jimenez:2014rna}. As an explanation for the large-angle anomalies (CMB), late-time anisotropies associated to imperfect dark energy can be a more natural starting point than primordial inflationary anisotropies, as the largest angles in the sky correspond to the present size of the universe and thus the scale of the present acceleration. In this case as well it is crucial to understand the general dynamical properties of anisotropic models in order to construct a potentially viable one in which the universe would end up in a late-time accelerating anisotropic attractor

The lay-out of the paper is as follows. In section \ref{set-up} we will briefly review the motivation for the models in the extra-dimensional D-brane scenarios and derive the general covariant equations for the gravity-scalar-vector system in four dimensions. In the following section \ref{cosmology} we will then adapt these equations to cosmology and present a very convenient formalism to analyse the rather complicated system in terms of bounded phase space variables. The closed dynamical system thus obtained is studied in section \ref{phase}, where we find the fixed points in two extreme (ultra-relativistic and weakly warped) regions and calculate the conditions for the existence and stability of each of the points. In section \ref{numerical} we show a few numerical solutions of the system. We then conclude in section \ref{conclusions}.

\section{The general set-up}\label{set-up}

We consider a gravitational system with a vector field $A_\mu$ and a scalar field $\phi$.  The main and novel feature is that the vector field lives in the metric $\hat{g}_{\mu\nu}$ that is {\it disformally} related to the gravitational metric $g_{\mu\nu}$. The relation is given as
\be \label{disformal}
\hat{g}_{\mu\nu}=C(\phi)g_{\mu\nu}+D(\phi)\phi_{,\mu}\phi_{,\nu} \,,
\ee
where $C(\phi)$ and $D(\phi)$ are the conformal and disformal factors, functions of the scalar field $\phi$, respectively. We use a comma for (simple or spacetime, which one is which is made clear by the context) partial derivatives, and our signature is $(-,+,+,+)$. The action we consider is then
\be \label{action}
S=\int d^4x \sqrt{-g}\lb \frac{R}{16\pi G} + p(\phi,X)\rb - \int d^4 x \sqrt{-\hat{g}}\lb \frac{1}{4}\hat{g}^{\al\bt}\hat{g}^{\ga\da}F_{\al\ga}F_{\bt\da} + \frac{1}{2}m^2 \hat{g}^{\mu\nu}A_\mu A_\nu\rb  \,.
\ee
The first piece includes standard Einstein gravity and the scalar with a Lagrangian that depends on the field and its kinetic term $X=-g^{\al\bt}\phi_{,\al}\phi_{,\bt}/2$. For a canonic scalar field one has $p=X-V$, where $V(\phi)$ is the scalar potential ($p$ is in fact the ``pressure'' of the field in fluid description). The second piece is a Proca Lagrangian for the vector which is coupled to the metric (\ref{disformal}). The field strength tensor is $F_{\al\bt}=A_{\bt,\al}-A_{\al,\bt}$ as usual.

This type of action can be for instance an effective four dimensional reduction of DBI brane scenarios in type IIA string theory: in that particular case the conformal and disformal factors are related as $C = (T_3 h)^{-1/2}$ and $D = (h/T_3)^\frac{1}{2}$, where $h(\phi)$ is the warp factor of the higher dimensional geometry with the dimensions of one per energy density and $T_3$ is the tension of the brane and has the dimension of energy density. The Lagrangian for the scalar field has the usual DBI form $p = (1-\ga^{-1})/h - V$, where $\ga = (1-2hX)^{-1/2}$ is the Lorentz factor for the movement of the brane. In the following we use the generalised definition
\be \label{lorentz}
\ga \equiv \lp 1-2 \frac{D}{C}X \rp^{-\frac{1}{2}}\,,
\ee
which encompasses, but is not limited to, the pure DBI example. As the branes are nothing but boundary conditions for open strings whose endpoints on the brane are described by a $U(1)$ field, taking into account vector field matter on the DBI brane seems indeed more natural than not to. Any kind of matter living on the brane will also be by construction disformally coupled to the DBI radion $\phi$ from our point of view, as that feels the induced metric on the brane. Commonly used geometries are the anti-de Sitter approximation $h \sim \phi^{-4}$ and the exact Klebanov-Strassler solution; the former can approximate regions sufficiently far from the ``tip of the throat'' of the spacetime. In the numerical examples below however, we will adopt a simple exponential ansatz for the warp factor as our present purpose is to explore the generic dynamical properties of these theories rather than investigate specific stringy embeddings of the models. 

Let us briefly comment on the possible origins of the vector field mass. In type IIB string theory, vector fields can acquire mass via the familiar Higgs mechanism or via a stringy St\"uckelberg mechanism that takes place when the vector couples to a two-form field in the 4D theory. Such couplings can be present in the Wess-Zumino action, though in some particular compactifications they are projected out of the spectrum by the action of orientifolds. In general we however have motivations to allow the vector field to have a mass in the action (\ref{action}), see \cite{Dimopoulos:2011pe,Koivisto:2013fta} for more detailed discussions.

We can write the action solely in terms of the two fields and the physical metric by eliminating the disformal metric using (\ref{disformal}). Upon employing the useful relations
\[
\tilde g^{\mu\nu} = \frac1C \left( g^{\mu\nu} + \ga^2 \frac{D}{C} \phi^{,\mu}\phi^{,\nu} \right) \,,
\]
and $\sqrt{-\tilde g} = \ga^{-1} C^2 \sqrt{-g}$, the vector portion of the action reads
\be
S_A = \int d^4 x \sqrt{-g} \lb \ga^{-1}\lp - \frac14 F^2 - \frac12 M^2A^2\rp + \frac12 \ga \h\phi^{,\al}\phi^{,\bt}\lp F_{\ga\al}F^{\ga}_{\phantom{\ga}\bt} 
+M^2A_\al A_\bt \rp \rb \,,
\ee
where $\h \equiv D/C$ and $M^2 \equiv Cm^2$. Note that here $\h$ is a generic notation for the ratio $\h=D/C$, but in a warped brane scenario one has specifically $C \sim D^{-1} \sim \h^{-\frac{1}{2}}$. 
We see that when the relation between the two metrics is purely conformal, $D=0$ and $\ga=1$, only the mass gets rescaled by $C$ as it breaks the conformal invariance of the Maxwell action. Generally, this action is suppressed by the Lorentz boost of the brane movement (\ref{lorentz}), and a new direction-dependent contribution appears proportional to gradients of the scalar field. We also note that the conformal part trivialises if the extra-dimensional geometry is (nearly) flat, whereas the disformal part could still be significant, and is in this sense more generic. 

The field equations that follow from the variation of the action (\ref{action}) are
\be
G_{\mu\nu} = \kappa^2\lp T^A_{\mu\nu} + T^{\phi}_{\mu\nu}\rp\,,
\ee
where the gravitational coupling constant is $\kappa^2 \equiv 1/8\pi G$; the stress energy tensor for the scalar field $\phi$ is
\be \label{step}
T^{\phi}_{\mu\nu} = \p g_{\mu\nu} + \p_{,X}\phi_{,\mu}\phi_{,\nu} \,,
\ee
while that of the vector is
\ba
\tg \, T^A_{\mu\nu} & = & F_{\mu\ga} F_\nu^{{\phantom{\ga}\ga}} + g_{\mu\nu} \lb \ga^{-2}\lp -\frac14 F^2 - \frac12 M^2A^2\rp + \frac12 \h \phi^{,\al}\phi^{,\bt} \lp F_{\ga\al}F^{\ga}_{\phantom{\ga}\bt} + M^2A_\al A_\bt \rp \rb \nonumber \\
& + & \h \lb \frac14 \phi_{,\mu}\phi_{,\nu} F^2 - \phi^{,\al}\phi^{,\bt} F_{\mu\al} F_{\nu\bt} + \phi_{,\ga}\phi^{,\ga} F_{\mu\al} F_{\nu}^{\phantom{\ga}\al} - \phi_{,\mu}\phi^{,\al} F_{\al\ga} F_{\nu}^{\phantom{\nu}\ga} - \phi_{,\nu}\phi^{,\al} F_{\al\ga} F_{\mu}^{\phantom{\ga}\ga} \rb \nonumber \\
& + & M^2 \lb \ga^{-2} A_\mu A_\nu + \frac12 \h\phi_{,\mu}\phi_{,\nu} A^2 + \h \phi^\ga A_\ga \lp\phi_{,\mu} A_\nu+\phi_{,\nu}A_\mu\rp \rb \nonumber \\
& + & \frac12 \ga^2 \h^2 \phi_{,\mu} \phi_{,\nu} \phi^{,\al}\phi^{,\bt} \lp F_{\ga\al}F^{\ga}_{\phantom{\ga}\bt} + M^2 A_\al A_\bt \rp \, ,
\ea
where it is convenient to define $\tg\equiv1/\ga$. Owing to the mutual coupling between the two fields, these two are not separately conserved, but, as it must, $\nabla_\mu(T_{\phi}^{\mu\nu}+T_A^{\mu\nu}) = 0$.

The disformally coupled theory appears rather complicated compared to the familiar Maxwellian vector theory. However, in concrete physical situations the effects of the coupling can remain quite intuitive. As an example, consider electromagnetic field residing on the brane. This set-up could be relevant for inflationary magnetogenesis \footnote{Identifying the vector field on the probe brane with our familiar Maxwell field could be possible, for example if by the end of inflation the moving brane collided with the stack of branes that can incorporate larger symmetry groups and our standard model.  The details of reheating and indeed obtaining the standard model are in general unsolved problems in string inflationary models and do not concern us too much here, our purpose a to set up the formalism dynamical analysis; we leave the construction of specific inflationary scenarios to further studies.}. We would then employ the ``gauge'' ansatz $A_\mu=(A_0, {\bar A}^T + {\bar \nabla} \chi)$ as a test field in flat Friedmann-Lema\^itre-Robertson-Walker (FLRW) geometry written in terms of conformal time $\tau$ as $ds^2=a^2(\tau)(-d\tau^2+d{\bar x}^2)$, where spatial 3-vectors are denoted by an overbar. We can remove the longitudinal scalar mode $\chi$ by a choosing a suitable gauge, and the temporal scalar mode $A_0$ turns out to be non-dynamical and decoupled. The transverse vector modes ${\bar A}^T$ associated to the physical photon particle turn out to have the action 
\be
S_A = \frac{1}{2}\int d\tau d^3x  \gamma\delta^{ij}\lp \frac{d{A}^T_i}{d\tau} \frac{d {A}^T_j}{d\tau} + \ga^{-2}A^T_i \bar{\nabla}^2 A^T_j\rp \,, 
\ee
where $\bar{\nabla}^2$ is the Laplacian. In Fourier space the equation of motion is thus
\be
\frac{d^2{{\bf A}}^T}{d\tau^2} + \gamma^{-1}\frac{d \gamma}{d\tau}\frac{d{{\bf A}}^T}{d\tau} -
\lp\frac{k}{\ga}\rp^2 {\bf A}^T=0\,. 
\ee
The effect of the disformal coupling is to suppress the propagation speed of perturbations by the Lorentz factor, and in addition there is a scale-independent friction term. It appears that such time-dependent sound speed would be relevant to magnetogenesis only in the case of decreasing $\gamma$, which in the brane world case would correspond to the brane moving away from the warped region.

In the following we study a more general case, taking into account in addition a possible mass of the vector and the backreaction of the vector to the spacetime geometry. This consequently means that we have to allow for anisotropic spacetimes.

\section{Cosmology}\label{cosmology}

In order to study the cosmology of the system we choose the metric to be of Bianchi I type, which describes anisotropic expansion. For simplicity we restrict to the axisymmetric case in which the metric can be parametrised as
\be
ds^2 = - dt^2 + e^{2\al(t)} \lb e^{2\sigma(t)} \lp dx^2 + dy^2 \rp + e^{-4\sigma(t)} dz^2 \rb \,.
\ee
Here $\al$ is the overall isotropic volume expansion (logarithm of the averaged scale factor), and $\sigma$ describes the anisotropy: the derivative of $\al$ gives the Hubble expansion and the derivative of $\sigma$ the shear. The cosmological equations become, an overdot denoting derivative with respect to $t$,
\ba
0 & = & \ddot{A} + (\dot{\al} + 4\dot{\sigma})\dot{A} + \ga^{-2} M^2A + \ga^2 \dot{A}\dot{\phi} \lp \h\ddot{\phi} + \frac12 \h'\dot{\phi}^2 \rp \,, \label{first} \\
0 & = & \lp \ddot{\phi} + 3\dot{\al}\dot{\phi} \rp p_{,X} + \ddot{\phi}\dot{\phi}^2 p_{,XX} + \dot{\phi}^2 p_{,\phi X} - p_{,\phi} \nonumber \\
& + & \frac12 \ga^3\e \dot{A}^2 \lb \h \lp \ga^2 \ddot{\phi} - (\dot{\alpha} + 4\dot{\sigma})\dot{\phi}\rp + \frac12 \ga^2 \h' \dot{\phi}^2 \rb \nonumber \\
& + & \frac12 \ga\e A^2 \lb \h M^2 \lp \ga^2\ddot{\phi} + (\dot{\alpha} + 4\dot{\sigma})\dot{\phi}\rp + \frac12 \ga^2 \h'M^2 \dot{\phi}^2 + 2 MM' \rb \,, \\
0 & = & 3\ddot{\sigma} + 9\dot{\al}\dot{\sigma} - \ga\e\dot{A}^2 + \ga^{-1}\e A^2 M^2 \,, \\
0 & = & 6\ddot{\al} + 9\dot{\al}^2 + 9 \dot{\sigma}^2 + 3 p + \frac12 \ga\e\dot{A}^2 - \frac12 \ga^{-1}\e A^2 M^2 \,, \label{fried2}\\
0 & = & 3\dot{\al}^2 - 3\dot{\sigma}^2 - p_{,X}\dot{\phi}^2 + p - \frac12 \ga^3\e\dot{A}^2 - \frac12 \ga\e A^2 M^2 \,. \label{last}
\ea
In this case the (inverse of the) Lorentz factor reads
\be
\tg = 1/\ga = \sqrt{1-\h\dot{\phi}^2} \,.
\ee

We convert all our equations using the e-folding number $\al$ as the time variable in the following.  Bear in mind that, since from this section onwards we do not need and employ any tensor equation, $\al$ always refers to the time-dependent e-folding number, not to be confused with a Lorentz indexl.  To set up a convenient phase space, we firstly completely specify our action by choosing it to be of the DBI type, that is: $$p(X,\phi) = \frac{1-\tg(X,\phi)}{\h(\phi)} - V(\phi)\,.$$  With this choice we now define the following variables for the scalar and vector
\be \label{variables}
x = \frac{1}{\sqrt{3\tg(\tg+1)}}\kappa\phi_{,\al} \,, \quad y = \sqrt{\frac{\kappa^2V}{3\dot{\al}^2}} \,, \quad
u = \frac{1}{\sqrt{6}}e^{-\al+2\sigma}\kappa A \,, \quad v = \frac{1}{\sqrt{6}\tg}e^{-\al+2\sigma}\kappa A_{,\al} \,, \quad  \cM = \frac{M}{\dot{\al}} \,.
\ee
Physically the pair $(x,y)$ represents the scalar field, while $(u,v)$ is associated with the vector; their $\tg$ normalisations are the ones we found make the equations look the most transparent. Notice that $\cM$ simply tells us how $M$ evolves compared to the overall expansion rate, and does not encode any direct dynamic effect upon the background geometry. The derivatives of the metric are described by the shear and the slow-roll parameter
\be
\Sa= \sigma_{,\al}\,, \quad \en = -\ddot{\al}/\dot{\al}^2 \,.
\ee
The Friedmann equation (\ref{last}) then beautifully\footnote{We can embellish further the Friedmann equation by defining $\tilde{u} = \frac{1}{\sqrt{6\tg}}e^{-\al+2\sigma}\kappa \cM A$ and $\tilde{v} = \frac{1}{\sqrt{6\tg^3}}e^{-\al+2\sigma}\kappa A_{,\al}$ so that $1-\Sa^2 = x^2 + y^2 + \tilde{u}^2 + \tilde{v}^2$, but we did not find this formulation any more convenient than the one we chose.} reads
\be \label{friedmann}
1-\Sa^2 = x^2 + y^2 + \cM^2 u^2/\tg + v^2/\tg \,,
\ee
and the second Friedmann equation (\ref{fried2}) constrains $\en$, almost as beautifully, as
\be
2\en = 3(\Sa^2+1) + 3\tg x^2 - 3y^2 - \tg \cM^2 u^2 + \tg v^2 \,.
\ee
Finally, to close the system we need to parameterise the slopes of the three $\phi$-dependent functions as
\be
\la_V = \sqrt{\frac32}\frac{V_{,\phi}}{\kappa V}\,, \quad \la_\h = \sqrt{\frac32} \frac{\h_{,\phi}}{\kappa \h}\,, \quad \lM = 2\sqrt{\frac32} \frac{M_{,\phi}}{\kappa M} \,.
\ee

We are then ready to write down the dynamical equations. These come out as follows:
\ba
2\tg^2 x_{,\al} & = & (1+2\tg)(1-\tg) \lb \SQRT \lK x^2 - \en x \rb + \Upsilon \,, \\
y_{,\al} & = & \SQRT \lV x y + \en y \,, \\
u_{,\al} & = & \lp 2\Sa - 1 \rp u + \tg v \,, \\
v_{,\al} & = & \lp \en - 2\Sa - 2 \rp v - \tg \cM^2 u \,, \\
\Sa_{,\al} & = & \lp \en - 3 \rp \Sa + 2 \tg v^2 - 2\tg \cM^2 u^2 \,, \\
\cM_{,\al} & = & \SQRT \lM \cM x + \en \cM \,,
\ea
where
\be \label{dlng}
\tg\tg_{,\al} = (\tg^2-1) \lb \SQRT \lK x - \en \rb + (\tg-1) \Upsilon /x \,;
\ee
notice that we have eliminated $\h$ thanks to $3x^2\h = \ga-1 = (1-\tg)/\tg$. The second derivative of the scalar field appears in the combination
\ba
\Upsilon & \equiv & \sqrt\frac{\tg+1}{3\tg} \kappa\phi_{,\al\al} \\
& = & \frac{\tg}{(\tg-1)\lb v^2 + \cM^2 u^2 \rb - \tg x^2} \left\{ \SQRT x^2 \lb \frac{1-\tg}{\tg} \lK \lp (1+2\tg) \tg x^2 + v^2 + \cM^2 u^2 \rp + 2 \tg^2 \lV y^2 + 2 \tg \lM \cM^2 u^2 \rb \right. \nonumber\\
&& \left. + (1+\tg) \lb 3\tg^2-\en \rb x^3 + \frac{1-\tg}{\tg} \lb v^2 \lp 4\tg^2\Sa^2 + \tg^2 + \en \rp +  \cM^2 u^2 \lp 4\tg^2\Sa^2 + \tg^2 - \en \rp \rb x \right\} \,, \nonumber
\ea
that is manifestly finite in the IR and UV limits.  We defined $\SQRT \equiv \sqrt{\tg(\tg+1)/2}$.

There are 5 distinct effective energy contributions to the Friedmann equation (\ref{friedmann}).  We choose to solve away $y$ assuming it is always positive \footnote{Negative energy density potentials $V<0$ could be allowed by extending the analysis to allow imaginary $y$.}, and the phase space is reduced to a four dimensional one.  In the following we shall consider the ultra-relativistic ($\tg=0$) and non-relativistic ($\tg=1$) regimes: one sees immediately from (\ref{dlng}) that $\tg$ is fixed at these limits; furthermore, the $\lK$ parameter also disappears from the system once $\tg$ is settled, as expected.

\section{The fixed points in phase space}\label{phase}

{\bf Lyapunov stability.} We identified the fixed points of the system and studied their stability by applying Lyapunov's analysis to the linearised system around each point.  For completeness, we recall the useful definitions and conditions for the fixed point $\vx_f$---we collect $\vx \equiv (x,y,u,v,\Sa,\cM)$.
\begin{itemize}
  \item $\vx_f$ is Lyapunov stable if, for every initial condition $\vx(0)$, there exists some $\ve>0$ such that $|\vx(t)| < \ve$ for all $t>0$. This is realised if and only if:
  \begin{itemize}
    \item[$\circ$] every eigenvalue $\Lambda_i$ of the linear system lies in the closed left half complex plane---$\Re(\Lambda_i) \leq 0$, and
    \item[$\circ$] every eigenvalue with zero real part is semisimple, that is, has coinciding algebraic and geometric multiplicity.
  \end{itemize}
  \item $\vx_f$ is semistable if $\lim_{t\rar\infty} \vx(t)$ exists for all initial conditions $\vx(0)$. This happens when and only when:
  \begin{itemize}
    \item[$\circ$] $\vx_f$ is Lyapunov stable, and
    \item[$\circ$] there is no nonzero imaginary eigenvalue---$\Im(\Lambda_i) = 0$.
  \end{itemize}
  \item $\vx_f$ is asymptotically stable (which we call simply ``stable'') if $\lim_{t\rar\infty} \vx(t) = \vx_f$ for all initial conditions $\vx(0)$. This is verified once and only once
  \begin{itemize}
    \item[$\circ$] every eigenvalue $\Lambda_i$ lies in the open left half complex plane---$\Re(\Lambda_i) < 0$.
  \end{itemize}
  \item $\vx_f$ is unstable if it is not Lyapunov stable.
\end{itemize}
We move on to classify the fixed points of the autonomous system in the two regimes of an ultra-relativistic and nonrelativistic brane.

\subsection{Ultra-relativistic regime}

In the limit $\tg \rar 0$ we obtain the following closed system
\ba
x_{,\al} & = & \frac12 \frac{x}{v^2+\cM^2u^2}\lb \lp 2\en+4\Sa+1 \rp v^2 + \lp 2\en-4\Sa-1 \rp \cM^2 u^2 \rb \,, \\
u_{,\al} & = & (2\Sa-1)u \,, \\
v_{,\al} & = & (\en-2\Sa-2)v \,, \\
\Sa_{,\al} & = & (\en-3)\Sa \,, \\
\cM_{,\al} & = & \en \cM \,.
\ea
First of all notice that there is no more reference to the three slopes $\lV$, $\lK$, and $\lM$: the system is independent on the structure of the scalar functions $C(\phi)$, $D(\phi)$, and $M(\phi)$---however, the stability properties of this solution depend on $\lK$ as it decides whether, upon smally perturbing the system, $\tg\rar0$ or not. The fixed points are: 1) de Sitter expansion with $y=\pm 1$ for any $\cM$ driven by the potential of the scalar field; 2) kinetic domination with $x=\pm 1$ (and everything else set to zero); 3) anisotropic solutions with $\Sigma = \pm 1$. The system ends up in one of these three stages whenever the warping increases without a bound. These asymptotic regimes themselves are not particularly interesting, as either the anisotropy is too large or it is completely diluted away by the scalar field.

\subsection{Nonrelativistic regime}

The nonrelativistic regime features much richer dynamics. Indeed, in the limit $\tg \rar 1$ we can cast the autonomous system as
\ba
x_{,\al} & = & (\en-3)x - \lV y^2 - \lM \cM^2 u^2 \,, \\
u_{,\al} & = & (2\Sa-1)u + v \,, \\
v_{,\al} & = & (\en-2\Sa-2)v - \cM^2 u \,, \\
\Sa_{,\al} & = & (\en-3)\Sa - 2 \cM^2 u^2 + 2 v^2 \,, \\
\cM_{,\al} & = & \lM \cM x + \en \cM \,.
\ea
Below we will report the properties of the fixed points. In addition to applying the Lyaponov method to the system of 5 variables we have checked whether $\tg=1$ remains stable in these cases.
In order to establish this we can perturb around the value $\tg=1$: it is then immediate to see that, since in the right hand side of (\ref{dlng}) the first term is linear in $(1-\tg)\geq0$ (there is no zeroth term), it suffices to study the sign of its coefficient:
\be\label{tgstab}
(1-\tg)_{,\al} \propto \lb (\lK + \lV) x^2 - 3x + (\lV - \lM) 2\cM^2 u^2 + \lV v^2 + \lV \Sa^2 - \lV \rb (1-\tg) \,.
\ee
For each solution $\vx_f$ we need to ensure that this coefficient is negative, which is where $\lK$, thus far apparently waned, comes back into the game.

\subsubsection{\scshape{Isotropic kinetic solution}}

\ba
  &\begin{dcases}
    x = \pm 1 \,,	& y = 0 \,, \\
    u = 0 \,,		& v = 0 \,, \\
    \Sa = 0 \,,		& \cM = 0 \,,
  \end{dcases} \\
  \centering
  &\en = 3 \,. \nonumber
\ea

{$\boldsymbol \triangleright$ }{\it Comments --} This solution is never stable, since one of the eigenvalues is +1.

\subsubsection{\scshape{Isotropic scalar domination solution 0}}
\label{solution1}

\ba
  &\begin{dcases}
    x = -\lV/3 \,,	& y = \sqrt{1-x^2} \,, \\
    u = 0 \,,		& v = 0 \,, \\
    \Sa = 0 \,,		& \cM = 0 \,,
  \end{dcases} \\
  \centering
  &\en = \lV^2/3 \,. \nonumber
\ea
\begin{center}
\begin{tabular}{|l|l|}
  \hline
  Existence	&  $|\lV| \leq 3$ \\ \hline
  Semistable	&  $-\sqrt6 \leq \lM \leq \lV \leq 0$ or $0 \leq \lV \leq \lM \leq \sqrt6$ \\ \hline
  		&  $\lM \leq -\sqrt6 \leq \lV \leq 0$ or $0 \leq \lV \leq \sqrt6 \leq \lM$ \\ \hline
  Stable	&  $\lM < -\sqrt6 < \lV < 0$ or $0 < \lV < \sqrt6 < \lM$ \\ \hline
  $\tg\rar1$	&  $(\lV+\lK)\lV>0$ \\
  \hline  
\end{tabular}
\end{center}

{$\boldsymbol \triangleright$ }{\it Comments --} This solution degenerates to the previous one when $\la_V^2=3$. It is a well-known accelerating solution for exponential quintessence. Now however its stability properties depend upon the embedding to the much more general model.

\subsubsection{\scshape{Isotropic scalar domination solution $\cM$1}}

\ba
  &\begin{dcases}
    x = 0 \,,		& y = 1 \,, \\
    u = 0 \,,		& v = 0 \,, \\
    \Sa = 0 \,,		& \cM = \text{any} \,,
  \end{dcases} \\
  \centering
  &\en = 0 \,. \nonumber
\ea
\begin{center}
\begin{tabular}{|l|l|}
  \hline
  Existence	&  $|\lV| = 0$ \\ \hline
  Lyapunov	&  $\Re\sqrt{1-4\cM^2} \leq 3$ \\ \hline
  Semistable	&  $|\cM| \leq 1/2$ \\ \hline
  $\tg\rar1$	&  always \\
  \hline  
\end{tabular}
\end{center}

{$\boldsymbol \triangleright$ }{\it Comments --} In this solution, the scalar sits at a minimum of its potential and drives the de Sitter expansion of the universe. The stability depends upon the ratio of the vector field mass and the de Sitter expansion rate.

\subsubsection{\scshape{Isotropic scalar domination solution $\cM$2}}

\ba
  &\begin{dcases}
    x = -\lV/3 \,,	& y = \sqrt{1-x^2} \,, \\
    u = 0 \,,		& v = 0 \,, \\
    \Sa = 0 \,,		& \cM = \text{any} \,,
  \end{dcases} \\
  \centering
  &\en = \lV^2/3 \,. \nonumber
\ea
\begin{center}
\begin{tabular}{|l|l|}
  \hline
  Existence	&  $\lM = \lV$ and $|\lV| \leq 3$ \\ \hline
  Lyapunov	&  $|\lV| \leq \sqrt6$ and $|\cM| \geq \sqrt{3\lV^2-18}/3$ \\ \hline
  		&  $\sqrt6 < |\lV| \leq 3$ \\ \hline
  Semistable	&  $0 \leq |\lV| < \sqrt3$ and $|\cM| \leq (3-\lV^2)/6$ \\ \hline
  		&  $\sqrt3 \leq |\lV| \leq 3$ and $\sqrt{3\lV^2-18}/3 \leq |\cM| \leq (\lV^2-3)/6$ \\ \hline
  $\tg\rar1$	&  $(\lV+\lK)\lV>0$ \\
  \hline  
\end{tabular}
\end{center}

{$\boldsymbol \triangleright$ }{\it Comments --} This is a generalisation of the solution \ref{solution1} for nonzero vector field masses.

\subsubsection{\scshape{Anisotropic solution 0}}
\label{solution0}

\ba
  &\begin{dcases}
    x = \pm\sqrt{1-\Sa^2} \,,	& y = 0 \,, \\
    u = 0 \,,			& v = 0 \,, \\
    \Sa = \text{any} \,,	& \cM = 0 \,,
  \end{dcases} \\
  \centering
  &\en = 3 \,. \nonumber
\ea
\begin{center}
\begin{tabular}{|l|l|}
  \hline
  Existence	&  $|\Sa| \leq 1$ \\ \hline
  Semistable	&  $\Sa=1/2$ and $\lV \leq \mp 3\sqrt2/2$ and $\lM \leq \mp 2\sqrt3$ \\ \hline
  $\tg\rar1$	&  $\lK x < 3$ \\
  \hline  
\end{tabular}
\end{center}

{$\boldsymbol \triangleright$ }{\it Comments --} As the shear of the metric and the kinetic energy of the scalar dilute both as stiff fluid energy densities, they can co-dominate the energy budget. However the solutions are semi-stable only at a specific point in the phase space when $\Sigma^2=1/4$ and $x^2=3/4$.

\subsubsection{\scshape{Anisotropic solution $\cM$}}

\ba
  &\begin{dcases}
    x = -3/\lV \,,		& y = 0 \,, \\
    u = 0 \,,			& v = 0 \,, \\
    \Sa = \sqrt{1-9/\lV^2} \,,	& \cM = \text{any} \,,
  \end{dcases} \\
  \centering
  &\en = 3 \,. \nonumber
\ea
\begin{center}
\begin{tabular}{|l|l|}
  \hline
  Existence	&  $|\lV| \geq 3$ and $\lM=\lV$ \\ \hline
  Lyapunov	&  $|\cM| \leq 1$ and $\sqrt{3-2\cM-\cM^2} \leq 1/6|\lV| \leq \sqrt{3+2\cM-\cM^2}$ \\ \hline
  Semistable	&  $|\cM| = 1 - 2\sqrt{1-9/\lV^2}$ \\ \hline
  $\tg\rar1$	&  $(\lV+\lK)\lV>0$ \\
  \hline  
\end{tabular}
\end{center}

{$\boldsymbol \triangleright$ }{\it Comments --} For these solutions also the universe expands as dominated by a stiff fluid.


\subsubsection{\scshape{Anisotropic vector solution A}}

\ba
 &\begin{dcases}
    x = \pm\sqrt{3/4} \,,	& y = 0 \,, \\
    u = \text{any} \,,		& v = 0 \,, \\
    \Sa = 1/2 \,,		& \cM = 0 \,,
  \end{dcases} \\
  \centering
  &\en = 3 \,. \nonumber
\ea

{$\boldsymbol \triangleright$ }{\it Comments --} This solution is never stable, since some eigenvalues are not semisimple. Note that it represents a variation of the solution \ref{solution0} allowing for nonzero $u$, which in turn nails $\Sa$ to $1/2$.

\subsubsection{\scshape{Anisotropic vector solution B}}

\ba
  &\begin{dcases}
    x = -\frac{6\lV}{\lV^2+12} \,,			& y = \frac{\sqrt{18(12-\lV^2)}}{\lV^2+12} \,, \\
    u = \pm\sqrt{\frac{6-\lV^2}{3(\lV^2-12)}} \,,	& v = \pm\frac{\sqrt{3(6-\lV^2)(\lV^2-12)}}{\lV^2+12} \,, \\
    \Sa = -2\frac{6-\lV^2}{\lV^2+12} \,,		& \cM = 0 \,,
  \end{dcases} \\
  \centering
  &\en = 6\lV^2/(\lV^2+12) \,. \nonumber
\ea

{$\boldsymbol \triangleright$ }{\it Comments --} This solution only exists for $\sqrt6 \leq \lV \leq 2\sqrt3$, but it is never stable, since some eigenvalues are not semisimple.

\subsubsection{\scshape{Anisotropic vector solution C}}

\ba
  &\begin{dcases}
    x = -\frac{\lV}{3}\frac{3u^2+1}{4u^2+1} \,,	& y = \frac{\sqrt{9-\lV^2+9u^2[9-\lV^2+2u^2(12-\lV^2)]}}{3(4u^2+1)} \,, \\
    u = \text{any} \,,				& v = \frac{u}{(4u^2+1)} \,, \\
    \Sa = \frac{2u^2}{(4u^2+1)} \,,		& \cM = \pm\frac{\sqrt{\lV^2-6+3u^2(\lV^2-12)}}{\sqrt3(4u^2+1)} \,,
  \end{dcases} \\
  \centering
  &\en = \lV^2(3u^2+1)/3(4u^2+1) \,. \nonumber
\ea
\begin{center}
\begin{tabular}{|l|l|}
  \hline
  Existence	&  $\lM=\lV$ and $\sqrt{\frac{6(4u^2+1)}{3u^2+1}} \leq |\lV| \leq 3\sqrt{\frac{24u^4+9u^2+1}{18u^4+9u^2+1}}$ \\ \hline
  Semistable	&  $|u| = \sqrt{\frac{\lV^2-6}{36-3\lV^2}}$ \\ \hline
  $\tg\rar1$	&  $(\lV+\lK)\lV>0$ \\
  \hline  
\end{tabular}
\end{center}

{$\boldsymbol \triangleright$ }{\it Comments --} For these solutions, the slow roll parameter satisfies $\epsilon=-\la_V^2(\Sigma-2)/6$. For an inflating universe we require $\epsilon$ close to zero, but using the minimum value of $\la_V$ we obtain $\epsilon=2$.

\section{Numerical solutions}\label{numerical}

In this section we will illustrate the properties of the solutions by plotting a few numerical examples.

\begin{figure}
\centering
\includegraphics[width=0.35\textwidth]{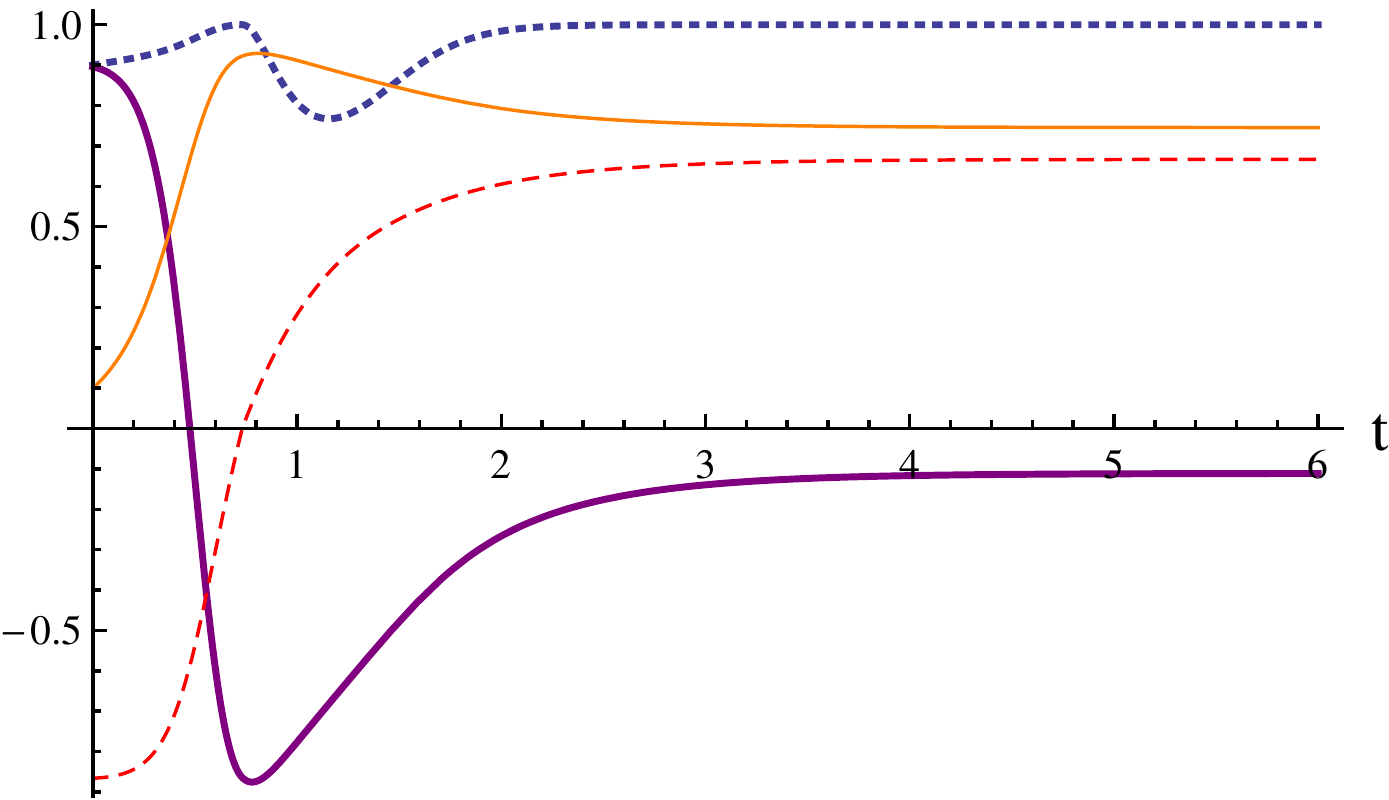}\hspace*{10pt}
\includegraphics[width=0.35\textwidth]{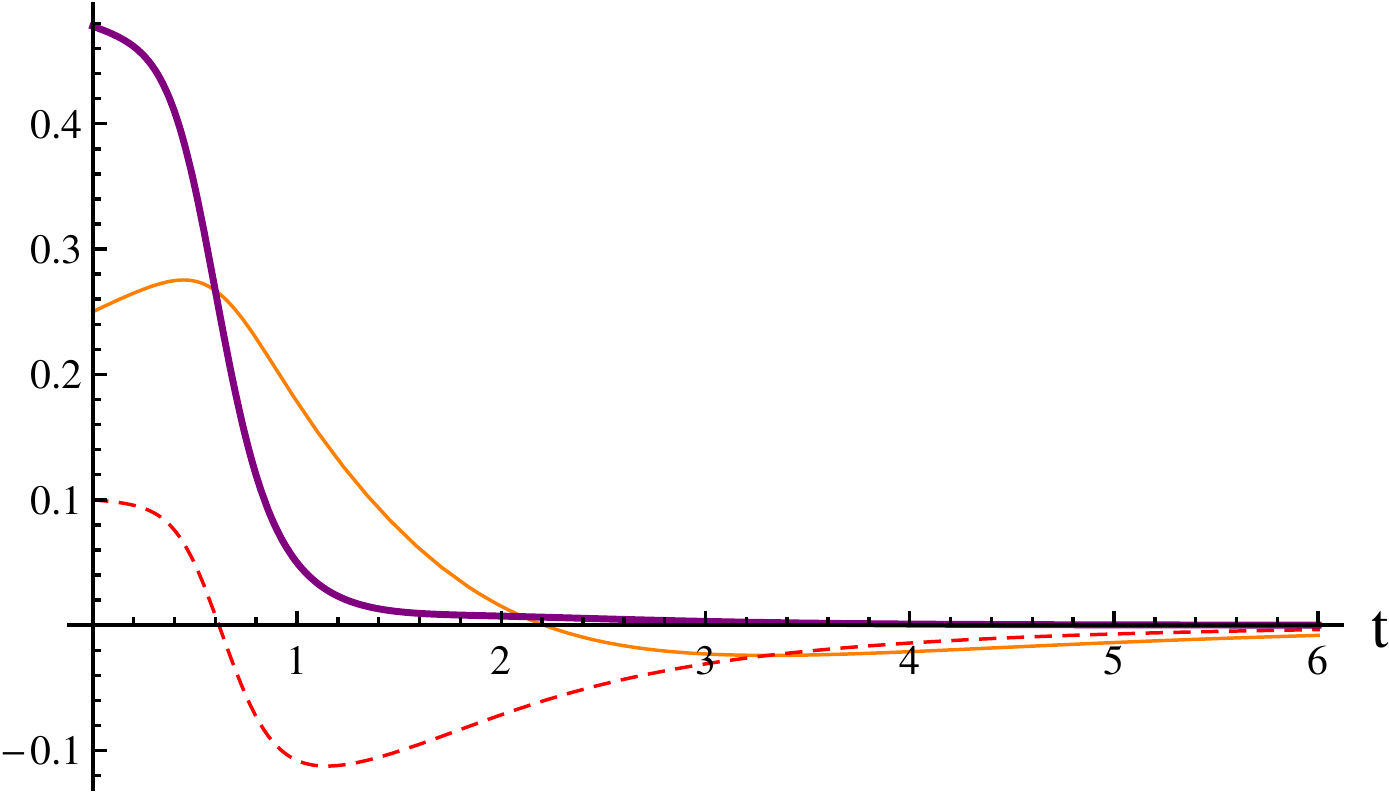}\hspace*{10pt}
\includegraphics[width=0.24\textwidth]{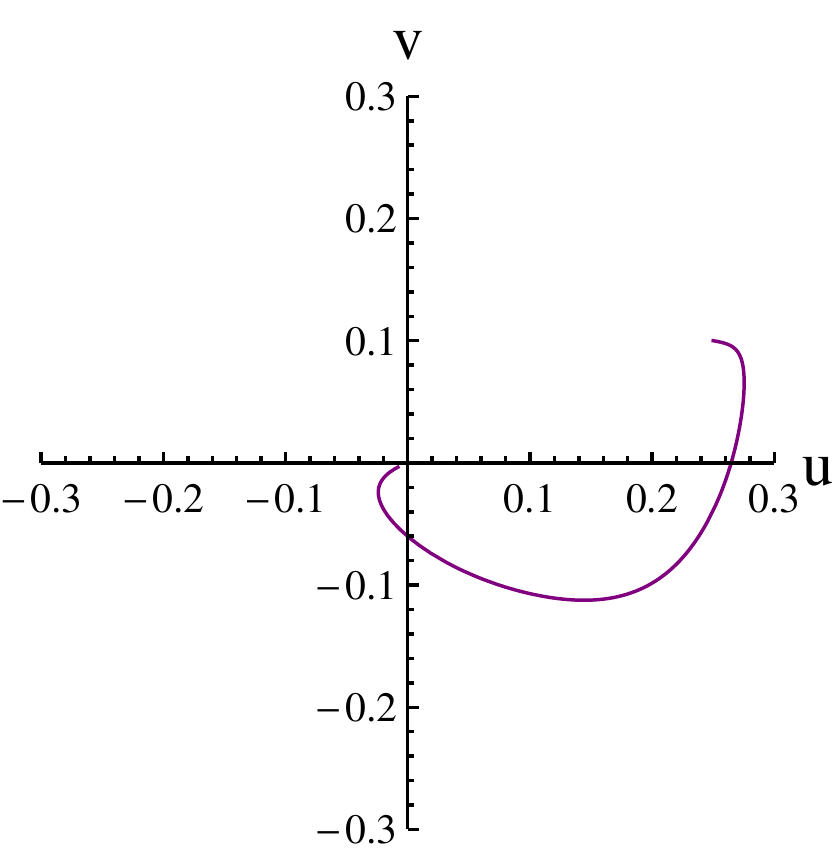}
\caption{The model $\lV=-2$ and $\lK=\lM=-3$, with $x_0=-\sqrt{3/4}$, $y_0=1/10$, $u_0=1/4$, $v_0=1/10$, $\cM_0=1/10$, and $\tg_0=9/10$.
{\bf Left panel}: The scalar field evolution. The dashed red line is $x$, the solid orange line is $y$. The $\tg$ is the thick dotted blue line and the thick purple line is the total equation of state.
{\bf Middle panel}: The vector field evolution. The dashed red line is $v$, the solid orange line is $v$. The shear $\Sa$ is the thick purple line.
{\bf Right panel}: The phase space evolution in the $(u,v)$ plane.}
\label{fig:model1}
\end{figure}

We begin with the first anisotropic vector solution A: this solution does not have any stable domain, so, as we see in Figure~\ref{fig:model1}, despite the initial conditions are quite close to the fixed point, it dynamically decays towards the isotropic solution 0, where it reaches instead a stable attractor.

\begin{figure}
\centering
\includegraphics[width=0.35\textwidth]{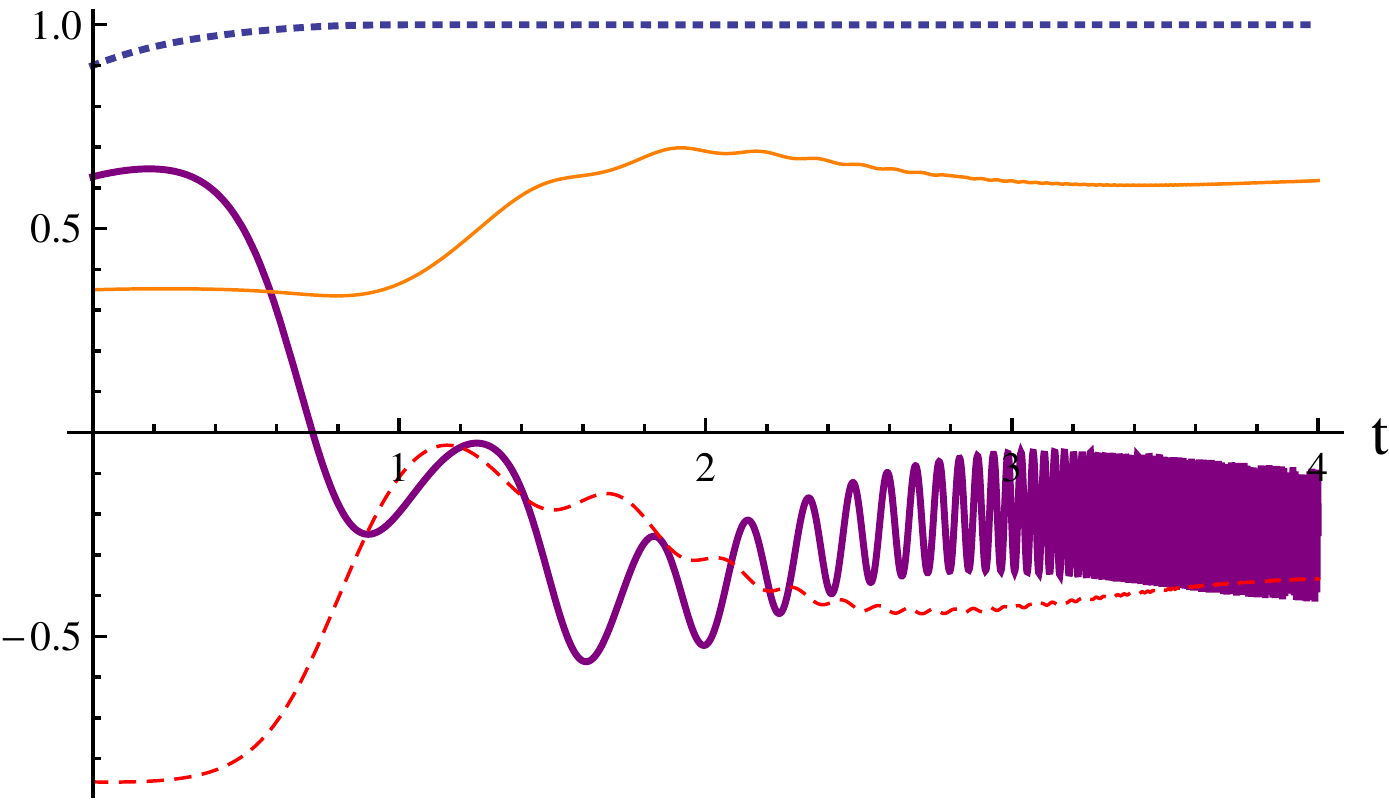}\hspace*{10pt}
\includegraphics[width=0.35\textwidth]{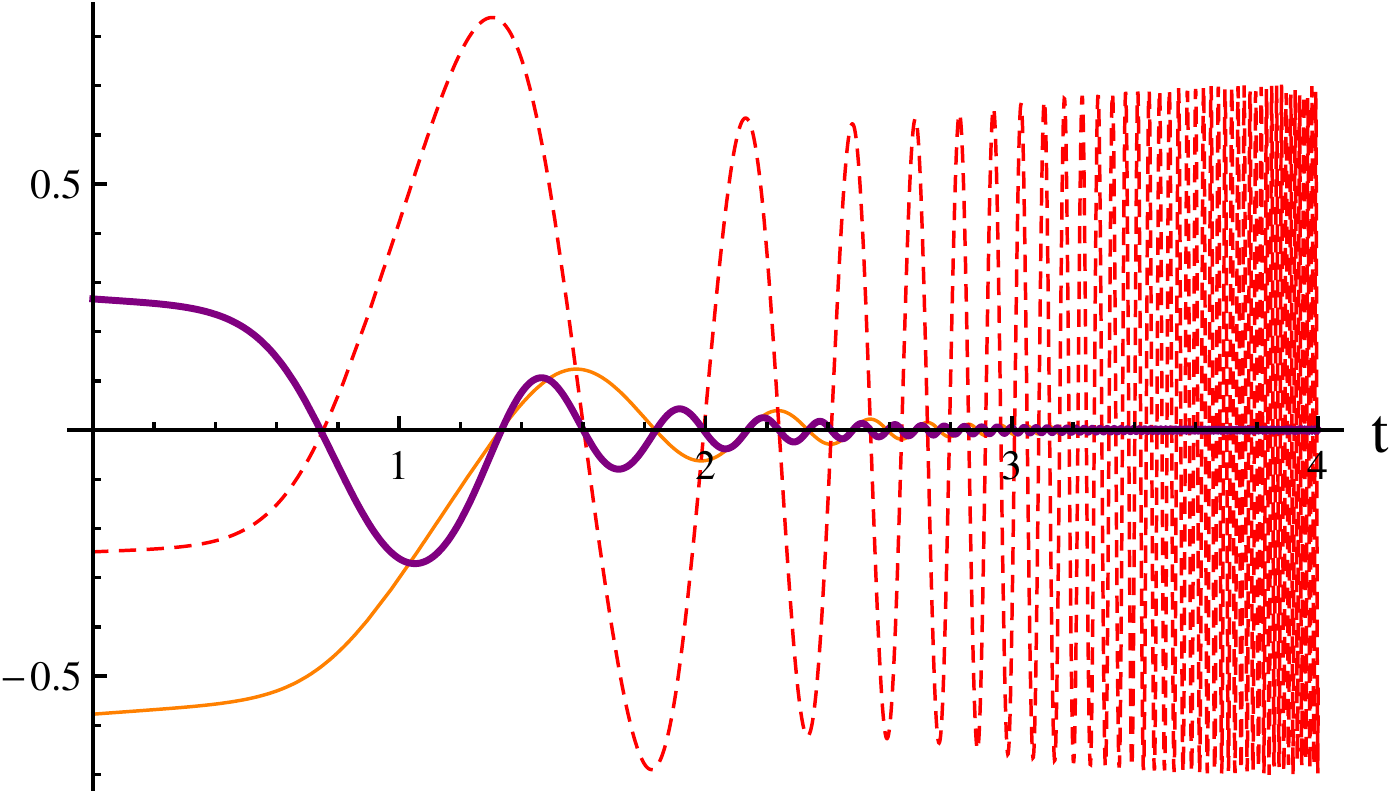}\hspace*{10pt}
\includegraphics[width=0.24\textwidth]{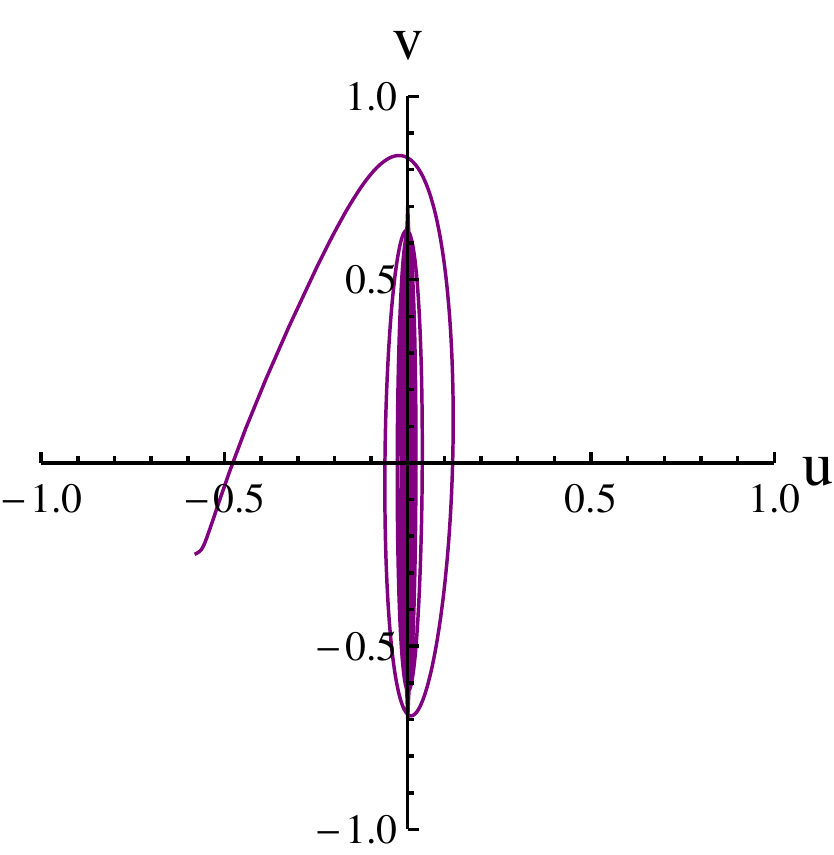}
\caption{The model $\lV=3$, $\lK=-1$, $\lM=-2$, with $x_0=-6/7$, $y_0=\sqrt6/7$, $u_0=-1/\sqrt3$, $v_0=-\sqrt3/7$, $\cM_0=1/10$, and $\tg_0=9/10$.
{\bf Left panel}: The scalar field evolution. The dashed red line is $x$, the solid orange line is $y$. The $\tg$ is the thick dotted blue line and the thick purple line is the total equation of state.
{\bf Middle panel}: The vector field evolution. The dashed red line is $v$, the solid orange line is $v$. The shear $\Sa$ is the thick purple line.
{\bf Right panel}: The phase space evolution in the $(u,v)$ plane.}
\label{fig:model2}
\end{figure}

In the next example we set $\lM=-2$ with $\lV=3$ and $\lK=-1$.  Once again, the initial conditions were chosen in order to start almost at the anisotropic fixed point B, but since this solution also is not stable, it will eventually decay, see Figure~\ref{fig:model2}.

\begin{figure}
\centering
\includegraphics[width=0.35\textwidth]{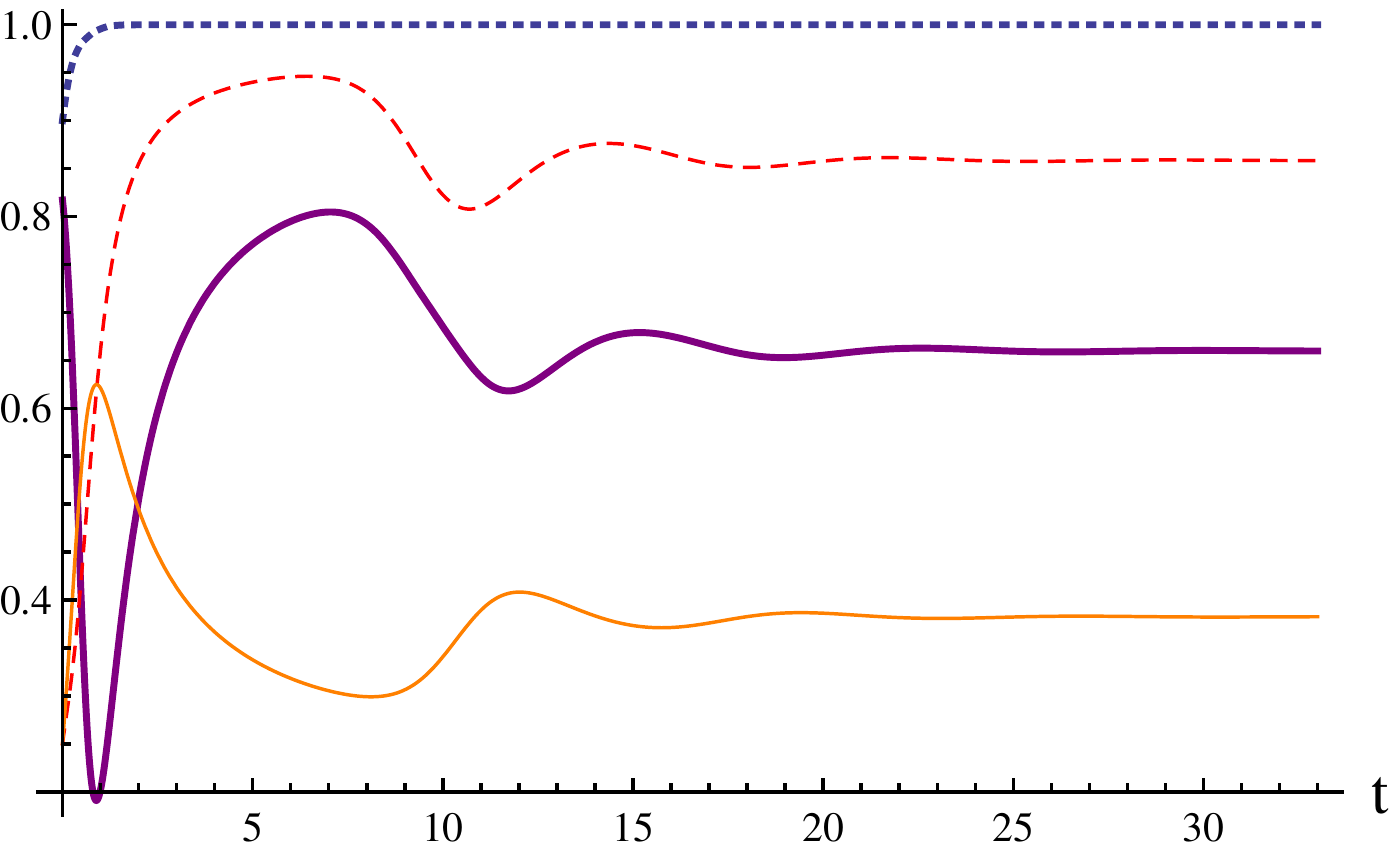}\hspace*{10pt}
\includegraphics[width=0.35\textwidth]{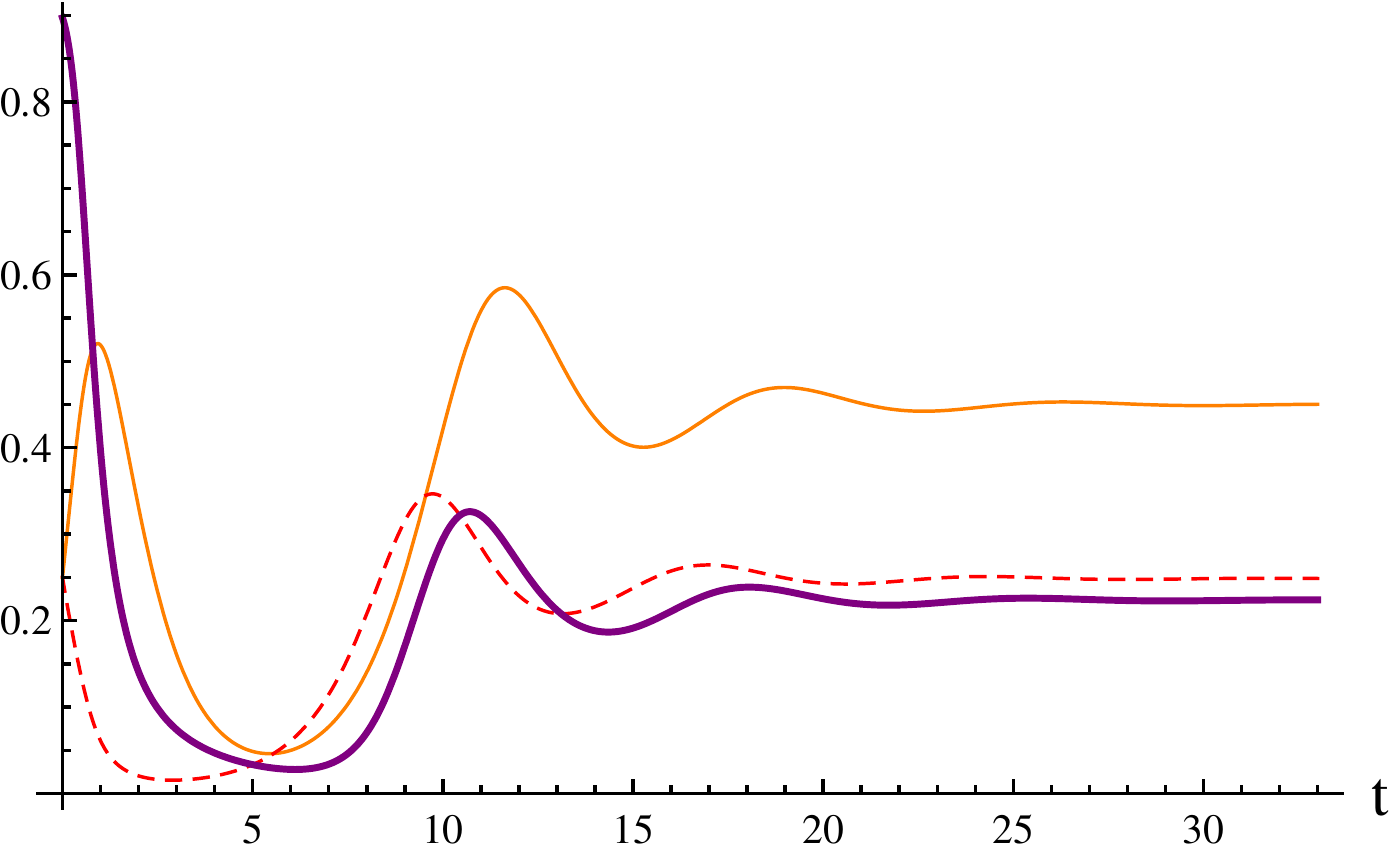}\hspace*{10pt}
\includegraphics[width=0.24\textwidth]{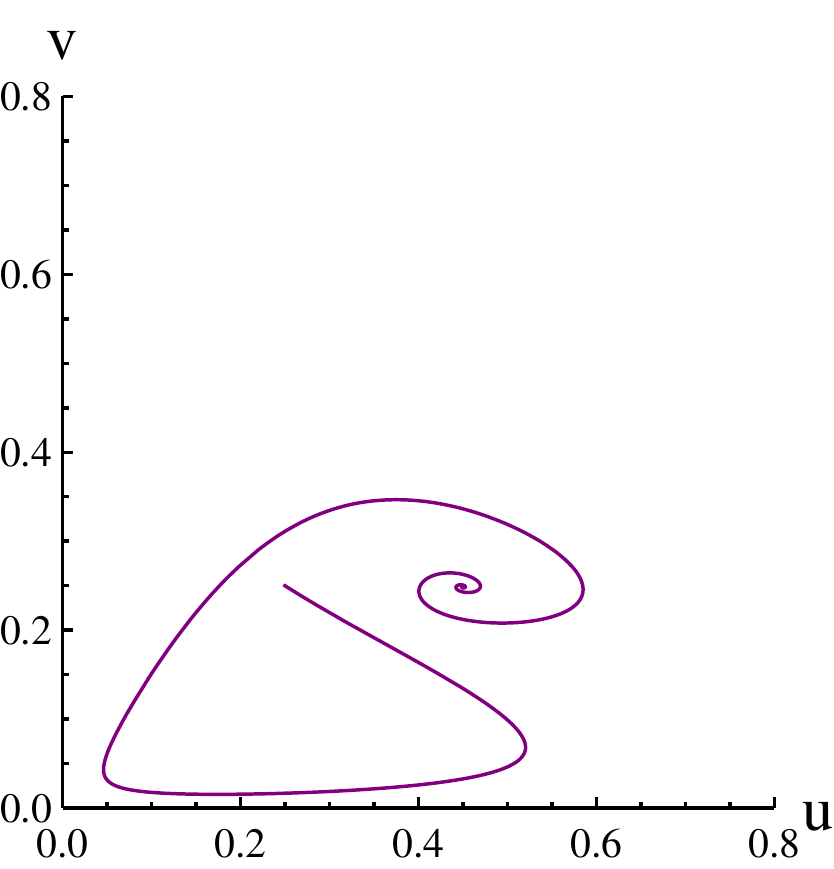}
\caption{The model $\lV=-2.9=\lM$, $\lK=-1$, with $x_0=y_0=u_0=v_0=1/4$, $\cM_0=1/10$, and $\tg_0=9/10$.
{\bf Left panel}: The scalar field evolution. The dashed red line is $x$, the solid orange line is $y$. The $\tg$ is the thick dotted blue line and the thick purple line is the total equation of state.
{\bf Middle panel}: The vector field evolution. The dashed red line is $v$, the solid orange line is $v$. The shear $\Sa$ is the thick purple line.
{\bf Right panel}: The phase space evolution in the $(u,v)$ plane.}
\label{fig:model3}
\end{figure}

In Figure~\ref{fig:model3} we instead show the case $\lV=-2.9=\lM$, $\lK=-1$, which is within the stable attractor domain of the anisotropic fixed point C.  Here the asymptotic future attractor will again be a non-relativistic ($\tg\rar1$) solution.  The vector will keep oscillating with decaying amplitude, so that neither $u$ nor $v$ are strictly constant.

\begin{figure}
\centering
\includegraphics[width=0.35\textwidth]{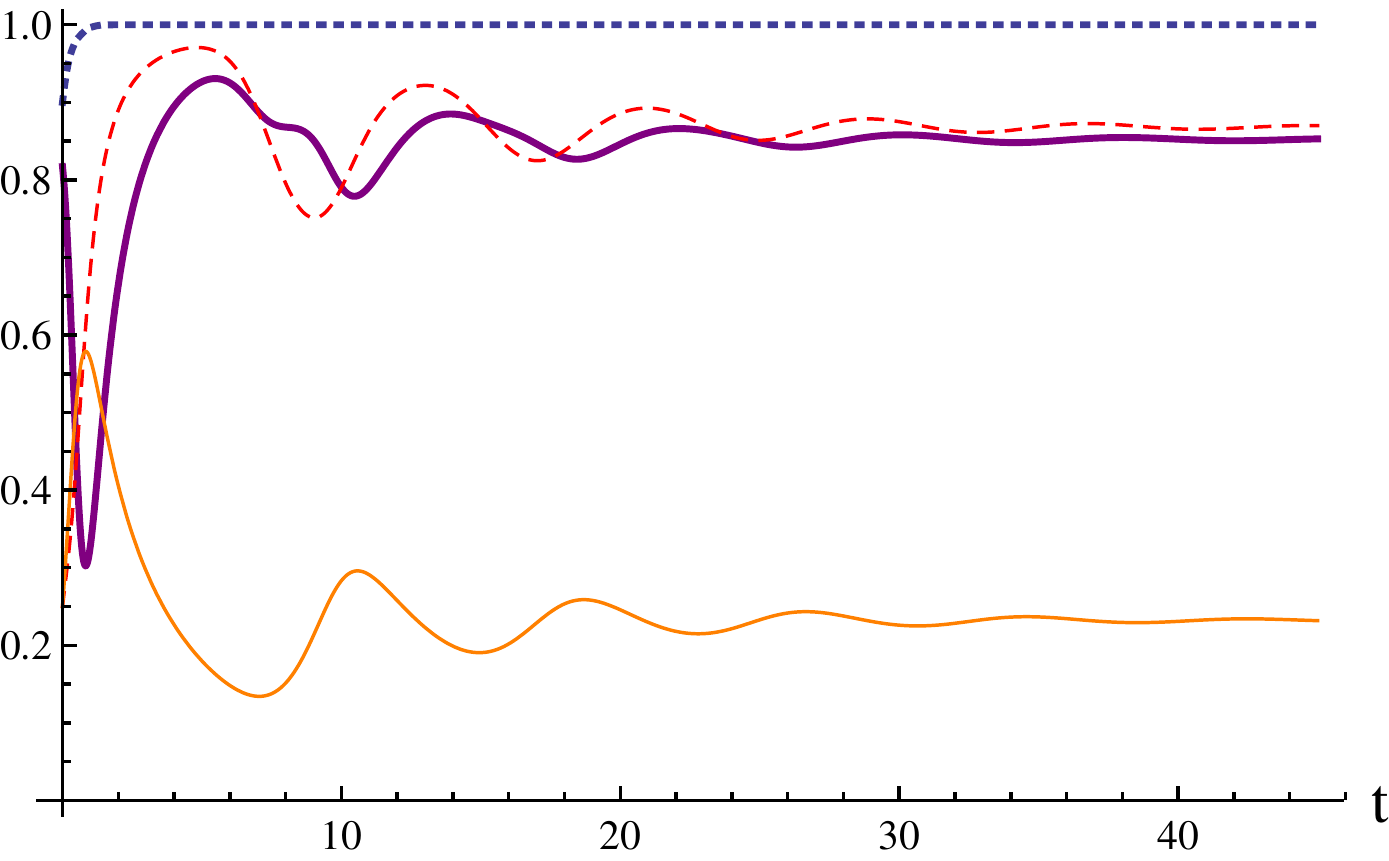}\hspace*{10pt}
\includegraphics[width=0.35\textwidth]{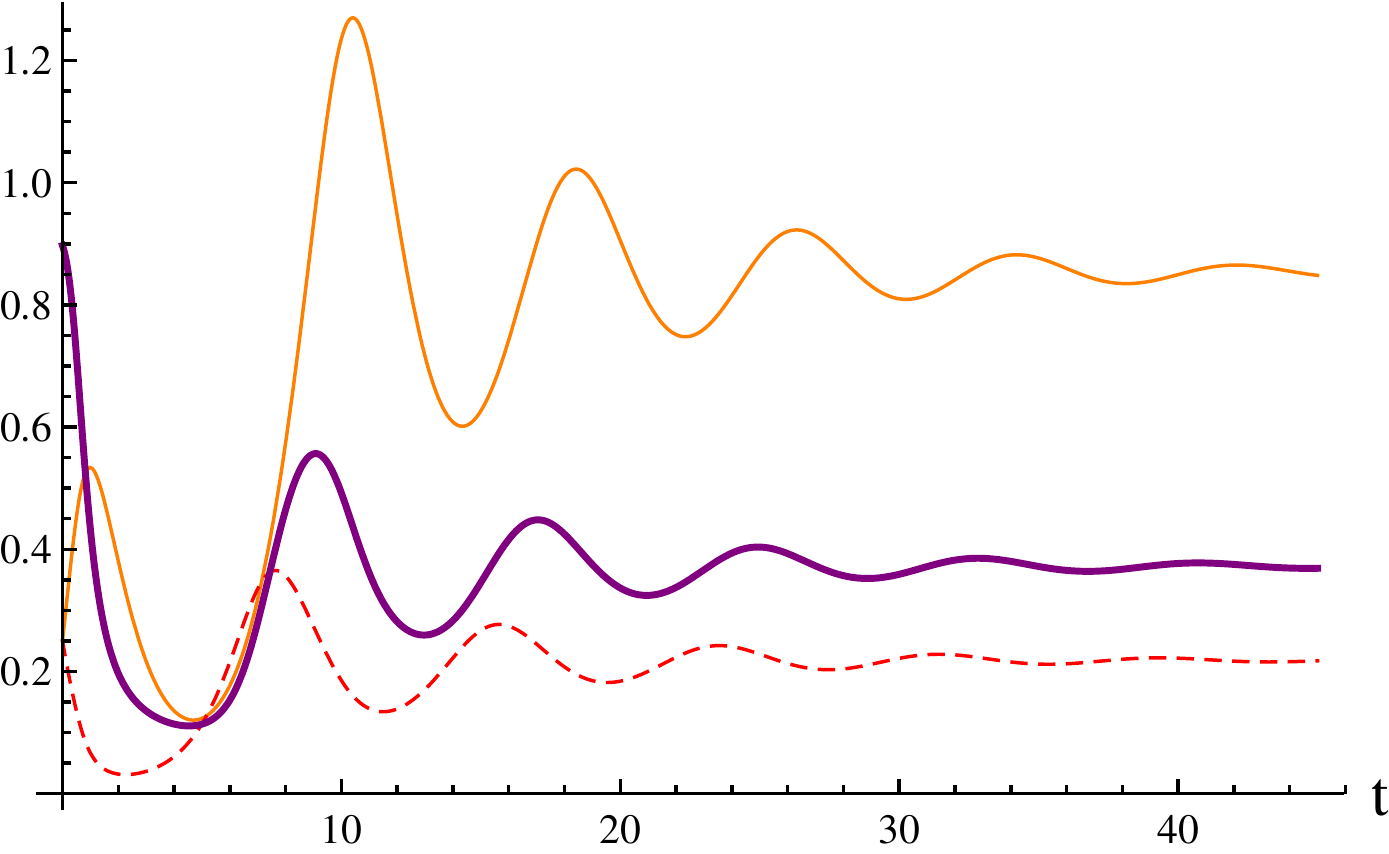}\hspace*{10pt}
\includegraphics[width=0.24\textwidth]{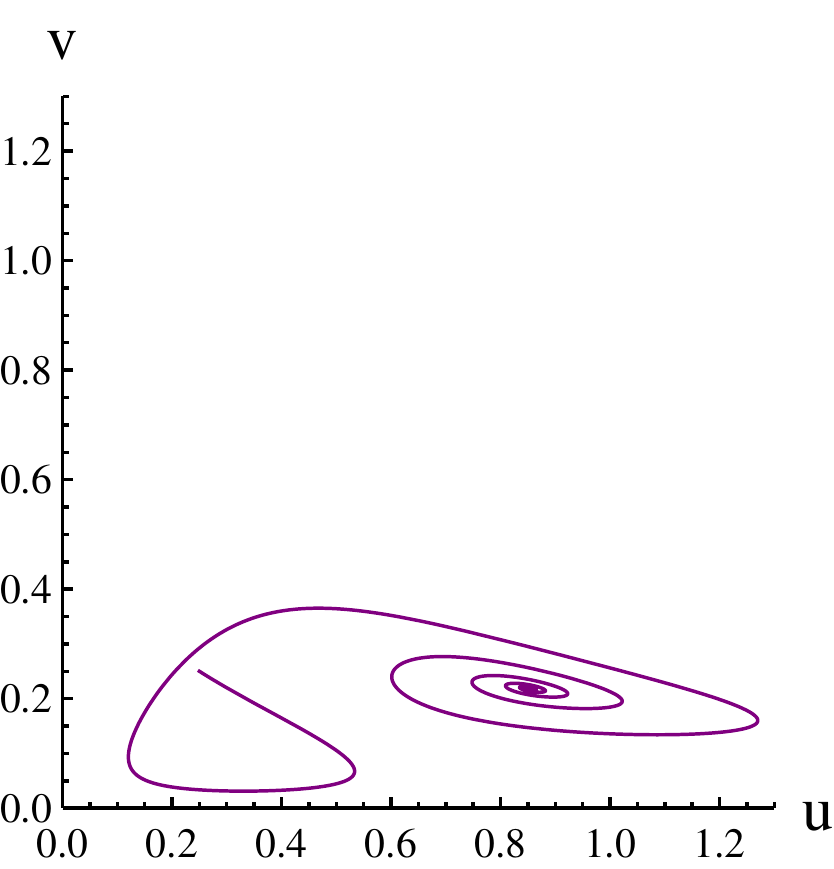}
\caption{The model $\lV=-3.2=\lM$, $\lK=-1$, with $x_0=y_0=u_0=v_0=1/4$, $\cM_0=1/10$, and $\tg_0=9/10$.
{\bf Left panel}: The scalar field evolution. The dashed red line is $x$, the solid orange line is $y$. The $\tg$ is the thick dotted blue line and the thick purple line is the total equation of state.
{\bf Middle panel}: The vector field evolution. The dashed red line is $v$, the solid orange line is $v$. The shear $\Sa$ is the thick purple line.
{\bf Right panel}: The phase space evolution in the $(u,v)$ plane.}
\label{fig:model4}
\end{figure}

Finally, in Figure~\ref{fig:model4} we plot a similar case where $\lV=-3.2=\lM$, $\lK=-1$.  The vector field and the shear both stabilise around non-zero values, leading to anisotropic expansion.  However, as is the case for all the solutions we obtained thus far, it is impossible to have slow-roll inflation and anisotropic hairs at the same time, since the slow-roll $\en$ parameter turns out to always be larger than one.

\section{Conclusions}\label{conclusions}

We studied the cosmological dynamics of disformally coupled vector fields.  We made a spatial ansatz for the vector field condensate and, in order to accommodate such choice, we picked the anisotropic Bianchi I spacetime, which includes anisotropic shear, as our cosmological metric.  We formulated the equations of motion for the system comprising a massive Abelian vector and a generic real scalar field in a very convenient autonomous form by carefully choosing the variables to sweep through the phase space.  The relatively cumbersome system becomes fairly manageable, and we were able to obtain several new fixed points, whose existence and stability properties we then studied systematically.

The models we focussed on are directly motivated by extra-dimensional scenarios in which the vector field resides on a moving brane.  The disformal coupling function is then uniquely given by the warp factor of the extra dimensional geometry, which in turn depends on the scalar field that describes the location of the brane---the scalar in this setup comes in a DBI form.  For concreteness, we assumed (nearly) exponential forms for both the warp factor and the scalar field potential, that is, we worked with, in first approximation, constant slopes $\lV$, $\lM$, and $\lK$.

For this particular model, our conclusion is that very generically the shear, which could in principle be supported by anisotropic stress of the spatial vector field, decays and the metric isotropises towards flat FLRW.  In spite of the fact that several anisotropic fixed points, including scaling solutions, exist in the phase space, they seem not to be of immediate cosmological relevance because either they are not attractors, or it is not possible to have simultaneously a slowly rolling vacuum energy driving the expansion, and small shear.

Nevertheless, the vector could still play a cosmological role even in the absence of shear. In particular, we confirmed in our more generic set-up the oscillating behaviour tha renders a vector field compatible with the FRW symmetries according to the cosmic vector isotropy theorem \cite{Cembranos:2012kk}. When the rapidly oscillating vector has, on average, a dust-like equation of state, such models can provide an alternative to dark matter in the form of disformally interacting massive particles a.k.a. DIMPs \cite{Koivisto:2012za,KOIVISTO:2013jwa}.

It should be stressed though that our conclusions strictly hold only for the very simplest possibility of constant or nearly constant parameters $\lV$, $\lM$, and $\lK$, whereas in general these will be functions of the DBI scalar field.  An analysis of more general field-dependent slopes and the construction of specific cosmological applications, such as models possibly suitable to generate inflationary magnetic fields or anisotropic hair, is left for future studies.

\subsection*{Acknowledgements}

We would like to thank Danielle Wills and Ivonne Zavala for useful discussions.  FU is supported by IISN project No.~4.4502.13 and Belgian Science Policy under IAP VII/37.


\bibliography{Drefs}

\end{document}